\documentclass[ preprintnumbers,A4paper, twocolumn,prb]{revtex4}
\usepackage{graphicx}
\usepackage{amssymb} %allows pretty double lined letters e.g. identity via \mathbb{I}
\usepackage[colorlinks=true, pdfstartview=FitV, linkcolor=red, citecolor=blue, urlcolor=blue]{hyperref}

\newcommand{\ket}[1]{| #1 \rangle}
\newcommand{\bra}[1]{\langle #1 |}

\begin{document}

\centerline {\Large \bf Efficient Cluster State  }
\smallskip
\centerline {\Large \bf Construction Under the }
\smallskip
\centerline {\Large \bf Barrett and Kok Scheme}
\smallskip
\smallskip
\centerline {\bf Simon Charles Benjamin}
\centerline {\small Department of Materials, University of Oxford.}
\centerline {\small s.benjamin@qubit.org}

\smallskip
\noindent {\bf 
Recently Barrett and Kok (BK) proposed an elegant method for entangling separated matter qubits. They outlined a strategy for using their entangling operation (EO) to build graph states, the resource for one-way quantum computing.  However by viewing their EO as a graph fusion event, one perceives that each successful event introduces an ideal redundant graph edge, which growth strategies should exploit. For example, if each EO succeeds with probability $p\gtrsim 0.4$ then a highly connected graph can be formed with an overhead of only about ten EO attempts per graph edge. The BK scheme then becomes competitive with the more elaborate entanglement procedures designed to permit $p$ to approach unity. 
}
\smallskip

In Ref.\ [\onlinecite{BKscheme}] Barrett and Kok (BK) describe a beautifully simple scheme for entangling separated matter qubits via an optical ``which-path-erasure'' process. Their scheme is necessarily probabilistic, with a destructive failure outcome that must occur at least 50\% of the time. Therefore they suggest using the process to construct a cluster state. The term cluster state, along with the more general term graph state, is used to refer to a multi-qubit entangled state with which one can subsequently perform `one-way' quantum computation purely via local measurements \cite{dan,jens}. The construction of such a state can tolerate an arbitrarily high failure rate, within the overall decoherence time, providing that successes are (a) known and (b) high fidelity. These properties are exhibited by the BK scheme and thus it is an efficient route to QC in the formal sense. However, in practice it is vital to know the overhead implied by the finite success probability $p$. 

In their paper Barrett and Kok suggest that it is necessary to build linear fragments of length greater than $1/p$ in order that, when one subsequently attempts to join those fragments onto some nascent cluster state, there will be a net positive growth. In fact the requirement seems to be rather less severe: the graph state created by a successful fusion of a simple EPR pair possesses a {\em redundant} `end' in such a way that when a subsequent addition fails, the total length does not decrease. (An EPR pair is equivalent, up to local unitaries, to an isolated two-qubit graph edge - I use term EPR in that sense.) A real decrease occurs only when a success is followed by {\em two consecutive} failures. 
 
\begin{figure}
  \begin{center}
    \leavevmode
\resizebox{6.7 cm}{!}{\includegraphics{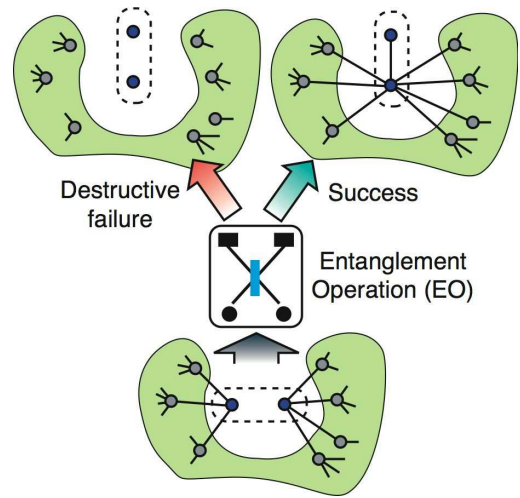}}
\end{center}
\caption{Effect of the BK scheme on arbitrary input cluster states. The outcomes are probabilistic: in ideal circumstances there is a 50\% chance of the success, yielding a form of fusion in the sense of Ref.\~[\onlinecite{dan}]. The set of connections radiating from the two marked qubits is completely arbitrary, including the case that there are no connections - in which case the process simply couples two isolated qubits to form a single graph segment (equivalent to an EPR pair).}
\label{fusion}
\end{figure}

\begin{figure}
  \begin{center}
    \leavevmode
\resizebox{6.5 cm}{!}{\includegraphics{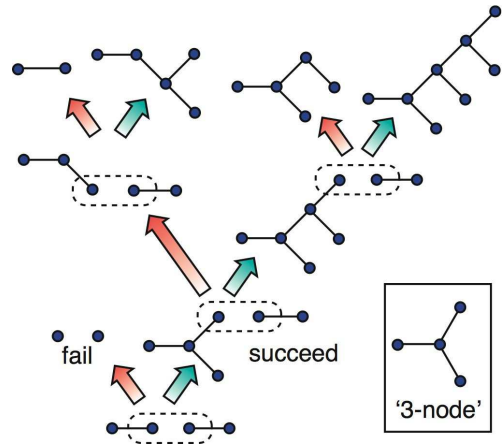}}
\end{center}
\caption{Simple strategy $\mathbb{S}1$ for growing a cluster state, here a linear chain. We use the rule in Fig. 1 and add in an EPR pair at each stage. In the uppermost row the fragments incorporate linear sections of 2,4,4,5 qubits respectively.  }

\label{linearBuild}
\end{figure}

\begin{figure}
  \begin{center}
    \leavevmode
\resizebox{6 cm}{!}{\includegraphics{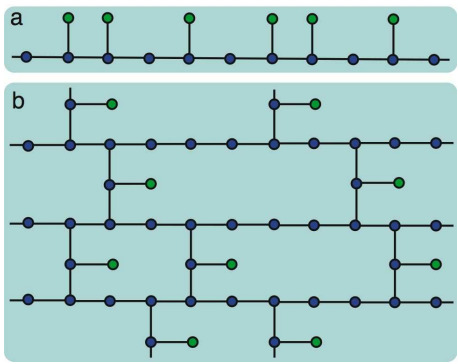}}
\end{center}
\caption{Above: a typical linear section growth by the methods depicted in Fig.\ (\ref{linearBuild}) will feature a large number of nodal `leaves' (green). Below: A higher dimensional network formed by fusing the nodal `leaves' of different linear sections, without supplying further EPR pairs. One successfully fuses a proportion $p$ of the original leaves, and these now provide a dense interconnected structure. There are further fusion opportunities afforded by the second generation leaves. Here the network is shown in 2D for clarity, but obviously the connections need not have this local structure. A graph of this kind is a resource for efficient one-way quantum computation. }
\label{network}
\end{figure}

The general action of a complete (two round) BK entanglement process is as shown in Fig. 1 (see Appendix for analysis) - it is evidently a fusion operation yielding a redundant qubit, in the sense of Ref. [\onlinecite{dan}]. The redundancy proves to be absolutely ideal for efficient cluster state construction, given the necessarily large failure probability. The simplest interesting strategy, which I denote $\mathbb{S}1$, is shown in Fig. 2: we prepare EPR pairs (at a cost $1/p$ operations each) and attempt to attach them to dangling bonds on the existing cluster. On success, the cluster gains two edges; on failure it loses one edge. The strict limit above which net growth is possible is then seen to be $p>1/3$. 

However, there are strategies involving preparing larger fragments prior to attachment to the main cluster, which perform better than $\mathbb{S}1$ regardless of $p$. (This is in contrast to the procedure in Ref.[\onlinecite{BKscheme}], where one only resorts to larger pre-prepared linear sections if growth is impossible otherwise.) The following strategy, $\mathbb{S}2$, is an example: 

\begin{enumerate}
\item \label{phase1} 
Prepare a 3-node from EPR pairs, as in Fig. 2. 
\item \label{nice2} Attempt to attach that 3-node to the main cluster;
\item \label{nice3} If successful, we have increased the number of edges by four - we may now go to (1) and repeat.
\item \label{nice4}  If unsuccessful, we have reduced the number of edges in the cluster by 1, and reduced our 3-node to a linear section of 3 qubits. We then attempt to upgrade this section back to a 3-node by attaching one further qubit to the central qubit. On success we jump to (2), on failure we have no remaining resources and must begin again at (1).
\end{enumerate}

This strategy will lead to cluster growth provided $p>{1\over 5}$. Over the full range ${1\over 2}> p> {1\over5}$ the strategy is less costly than $\mathbb{S}1$, an observation which suggests that in general the optimal strategy may involve growing large fragments prior to attachment. At $p={1\over 2}$, the cost per edge is $6$, compared to $14$ for the BK scheme according to the quantity $C_3=(p^{-2}+p^{-1}+1)/(3p-1)$. At $p=0.4$ the cost per edge is $13$, which compares to $48.75$ under BK - a trend to increasing gain as $p$ falls. When one introduces recycling into the BK strategy (as they suggest) then their costs do fall slightly: $12$ for $p={1\over 2}$ and $41.25$ for $p=0.4$ - but the same observations apply.

As shown in Fig. 2 and the upper part of Fig. 3, when one creates a linear section using $\mathbb{S}1$ or $\mathbb{S}2$ there will be a large number of apparently redundant nodal `leaves'. These could of course be pruned off by Z-measurements, once the target length is obtained. However they are in fact highly useful: they permit one to join together linear sections into higher dimensional arrays {\em without} additional EPR fragments. This is indicated in the lower part of Fig. 3. We will successfully convert a proportion $p$ of our leaves to the `T' cross pieces. At $p=0.4$, our cost-per-edge for building quasi-linear sections under $\mathbb{S}2$ was $13$; we lose only small proportion of our total number of edges as we connect these linear sections, and I estimate the final cost at about $16$. I emphasize that there is no reason to suppose that strategy $\mathbb{S}2$ is optimal. The potential efficiency of this BK based approach is therefore comparable to non-destructive growth schemes such as the recently proposed ingenious ``repeat-until-success'' process [\onlinecite{almut}].

I have recently been made aware\cite{munro} that the BK scheme may also be enhanced at a another level, in parallel to the strategy refinements explained here. The idea is to address the steps involved {\em within} each EO. The BK protocol involves a clever `double heralding' which filters out the unwanted component $\ket{11}$ from the qubit state, even when photon loss is present. As a development of this idea, one can postpone the filtering steps from successive EO's and subsume them into a single subsequent step.

Thanks to Sean Barrett, Dan Browne, Earl Campbell, Jens Eisert, Pieter Kok, Bill Munro and Tom Stace for helpful discussions.
\bigskip

\noindent {\bf Appendix: Analysis of the Fusion Process }
\smallskip

{\small 
The state specified by the
`before' diagram in Fig. 1 can be written as:
\[
\left(\ket{0}+\ket{1}\sigma^Z_{\rm L}\right)\left(\ket{0}+\ket{1}\sigma^Z_{\rm R}\right)\ket{X}
\]

Here $\ket X$ represents the graph state obtained by deleting the two marked qubits. The operator $\sigma^Z_{\rm L}\equiv \prod  \sigma^Z_1\sigma^Z_2...\sigma^Z_j$ is the product of $\sigma^Z$ operators applied to each of those qubits $1..j$ inside $\ket{X}$ to which our {\em left} hand qubit is attached by a graph edge. The operator $\sigma_R^Z$ is analogously defined for our right hand qubit. Following the BK procedure, prior to measurement we make a $\sigma^X$ operation on (say) the left qubit.

\[
\left(\ket{10}+\ket{11}\sigma^Z_{\rm R}+\ket{00}\sigma^Z_{\rm L}+\ket{01}\sigma^Z_{\rm L}\sigma^Z_{\rm R}\right)\ket{X}
\]

 The action of the optical entanglement process is defined by one of the four projection operators, 
 $\ket{00}\bra{00}$, $\ket{11}\bra{11}$, $\ket{10}\bra{10}\pm\ket{01}\bra{01}$. Each is associated with a unique measurement signature. The former two are the destructive failures, and the latter two are the successes. Assuming success we have 
\[
\left(\ket{10}\pm\ket{01}\sigma^Z_{\rm L}\sigma^Z_{\rm R}\right)\ket{X}
\]

 Now we flip the left qubit again, and fix the minus sign, if it has occurred, with a $\sigma_Z$ on either qubit.
  
\[
\left(\ket{00}+\ket{11}\sigma^Z_{\rm L}\sigma^Z_{\rm R}\right)\ket{X}
\]

Evidently this is a state where a single redundantly encoded qubit (in the sense of Ref.\~[\onlinecite{dan}]) inherits all the bonds of the previous pair (modulo 2, i.e. if the previous pair bonded to a common qubit in $\ket{X}$, there is no bound to that qubit).

The final cluster state in Fig. 1 is then obtained simply by applying a Hadamard rotation to one of our pair, which now becomes our `leaf'. Obviously these steps subsequent to the measurement can be compressed to a single operation on one qubit.
}

\end{document}